\documentclass[prl,twocolumn, showpacs,superscriptaddress]{revtex4}

\usepackage{amsfonts}
\usepackage{amssymb}
\usepackage[dvips]{graphicx}
\usepackage{color}
\usepackage{amsmath}

\newcommand{\bra}[1]    {\langle #1|}
\newcommand{\ket}[1]    {| #1 \rangle}

\begin{document}

\title{Entanglement Percolation with Bipartite Mixed States}

\author{S.~Broadfoot}
\affiliation{Clarendon Laboratory, University of Oxford, Parks Road, Oxford OX1 3PU, United Kingdom, EU}
\author{U.~Dorner}
\affiliation{Clarendon Laboratory, University of Oxford, Parks Road, Oxford OX1 3PU, United Kingdom, EU}
\author{D.~Jaksch}
\affiliation{Clarendon Laboratory, University of Oxford, Parks Road, Oxford OX1 3PU, United Kingdom, EU}
\affiliation{Centre for Quantum Technologies, National University of Singapore, Singapore, 117543}

\date{\today}

\begin{abstract}
  We develop a concept of entanglement percolation for long-distance
  singlet generation in quantum networks with neighboring nodes
  connected by partially entangled bipartite mixed states. We give a
  necessary and sufficient condition on the class of mixed network
  states for the generation of singlets. States beyond this class are
  insufficient for entanglement percolation.  We find that neighboring
  nodes are required to be connected by multiple partially entangled
  states and devise a rich variety of distillation protocols for the
  conversion of these states into singlets. These distillation
  protocols are suitable for a variety of network geometries and have
  a sufficiently high success probability even for significantly
  impure states. In addition to this, we discuss possible further
  improvements achievable by using quantum strategies including
  generalized forms of entanglement swapping.
  \end{abstract}

\maketitle

The distribution of entanglement through quantum networks is of
essential importance to quantum cryptography and distributed quantum
computing. However, the generation of entanglement between remote
nodes in a network faces a severe obstacle. Due to noise, e.g. in
transmission lines, desired maximally entangled states will degrade
into mixtures and the re-establishment of high-fidelity entanglement
requires sophisticated purification schemes, e.g. involving quantum
repeaters~\cite{Briegel98,Duan01}.  Although quantum repeaters are a
promising tool for quantum communication they still require a
considerable overhead of physical resources or are relatively
slow~\cite{Duer99}. An alternative scheme was recently proposed by
Ac\'in et al.~\cite{Acin07} in which ideas from classical bond
percolation have been applied to lattice-shaped quantum networks. It
was shown that maximally entangled singlet states can be created
between arbitrary points of the network, with a probability that is
independent of the distance between them, if the network nodes were
initially connected by partially entangled {\em pure} states with
sufficiently high entanglement.  The scheme relies on the fact that
states of this type can be converted by local operations and classical
communication (LOCC) into singlet states with finite
probability~\cite{Vidal99}. If this singlet conversion probability
(SCP) exceeds a certain, lattice-geometry-dependent threshold,
arbitrarily large clusters of network nodes connected by singlets are
created.  By successively applying entanglement swapping it is then
possible to establish singlets between arbitrarily distant nodes in a
cluster.  Although being a very promising concept, this classical
entanglement percolation (CEP) is not optimal. Indeed it was shown
in~\cite{Acin07,Perseguers08,Lapeyre09,cuquet09,perseguers09b}
that quantum strategies can be used to improve the SCP leading to even
more powerful protocols.

\begin{figure}[ht]
\centering\includegraphics[width=8.4cm]{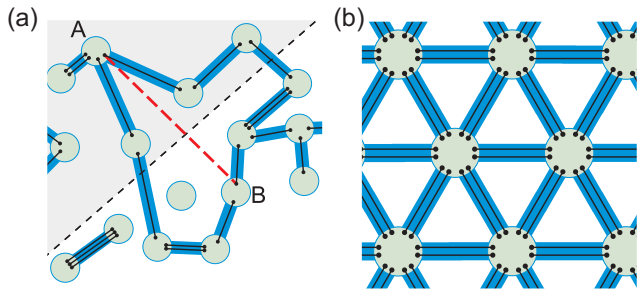}
\caption{%
  (color online) (a) General mixed-state quantum network. Qubits
  (black dots) in nodes (circles) may be connected by bonds (thick,
  blue lines). Each bond consists of one or more mixed entangled
  states, i.e. edges (black, solid lines), connecting two qubits in
  different nodes. The red, dashed line indicates a singlet which we
  attempt to create connecting nodes $A$ and $B$. The grey-shaded,
  upper-left region and the white, lower-right region contains nodes
  of groups $\mathcal{A}$ and $\mathcal{B}$, respectively (see text).
  (b) Example of a triangular lattice with two edges per bond: each qubit
  in a node is part of an entangled 2-qubit state.  As in (a),
  entanglement is represented by solid, black edges and a group of
  edges connecting two nodes form a bond. }
\label{fig.1}
\end{figure}

CEP, particularly when extended by quantum strategies, offers the
possibility to establish entanglement in quantum networks over long
distances. However, in previous work this has only been considered for
pure states.  In this paper we develop a concept of entanglement
percolation for mixed states. We start by analyzing the minimum
requirements for singlet generation in quantum networks where the
nodes are initially connected by a finite number of mixed 2-qubit states.  We show that the generation of a singlet
between two arbitrary nodes is possible if and only if they are
connected by at least two `paths' consisting of a particular class of
mixed states [see fig.~\ref{fig.1} (a)]. If this is not the case then
a singlet in such a network can never be formed between the nodes using LOCC.
Fortunately, it turns out that these mixed states are not only of
theoretical interest but arise naturally in systems undergoing
amplitude damping which is relevant for many practical setups, e.g.
where photon loss or spontaneous photon emission is dominant. Note
that, as in previous work on entanglement percolation~\cite{Acin07},
the sole requirement we impose is that the network initially contains
only bipartite entangled qubits. Multipartite entanglement might allow
for more general percolation schemes. We then specialize on regular
networks [see fig.~\ref{fig.1}(b)], present protocols for the
conversion of such mixed states into singlets and determine their SCP
revealing that CEP is possible with mixed states requiring only a
moderate overhead of qubits compared to CEP with pure states.
Moreover, the states can be substantially mixed. For instance, a
minimum purity of $72\%$ is required in a triangular lattice where
neighboring nodes are connected by two mixed states ($57\%$ for three
states). Furthermore, going beyond CEP, we develop additional
strategies which can further increase the success probability of
singlet creation and thus the prospect of efficient long distance
entanglement distribution.

 We start by proving a necessary and sufficient condition for
  a perfect singlet to be generated from an arbitrary network composed
  of 2-qubit states.  Throughout this paper we assume that gates and
  measurements are perfect (although this assumption is not required
  for the necessary condition). It is a first step to reveal the
  possibilities of entanglement percolation in mixed-state networks.
A general (mixed-state) quantum network is schematically shown in
fig.~\ref{fig.1}(a). The nodes, each of which can consist of several
qubits, may be connected by {\em bonds} consisting of a finite number
of {\em edges}. Each edge connects two qubits and represents a
partially entangled bipartite mixed state, i.e.  each qubit is
connected by exactly one edge to another qubit. We aim to identify a
condition for the state of the network for successfully creating a
singlet between two arbitrary nodes, say $A$ and $B$ via LOCC. To this
end we consider a partition of the network in two groups of nodes (or
qubits), a finite group $\mathcal{A}$ containing $A$ and a
group $\mathcal{B}$ consisting of the rest of the network, which
includes $B$ [see shaded areas in fig.~\ref{fig.1}(a)].
These groups are linked by a finite number of edges. In
  Ref.~\cite{Jane02} it was proven that a singlet between the groups
can be established with finite probability via operations local on
each group and classical communication if and only if $\mathcal{A}$
and $\mathcal{B}$ are connected by at least two mixed states which -
up to local unitaries - have the form
\begin{equation}
\rho(\alpha,\gamma,\lambda) = \lambda \ket{\psi}\bra{\psi} + (1-\lambda)\ket{01}\bra{01},
\label{eq:state1}
\end{equation}
where $\ket{\psi} = \sqrt{\alpha} \ket{00} +
\sqrt{1-\alpha-\gamma}\ket{11} + \sqrt{\gamma}\ket{01}$ and
$0\le\lambda\le1$. States like this occur naturally if a state
$\ket{\psi}$ is subject to amplitude damping. We distill a singlet
from two states of this type, $\rho(\alpha,\gamma,\lambda)$ and
$\rho(\beta,\delta,\nu)$, with a finite probability, by performing a
C-NOT gate between the qubits in each group and measuring the two
target qubits, i.e. the two qubits originally forming the state
$\rho(\beta,\delta,\nu)$, in the computational basis. If in both cases
the qubit is found in the state $\ket{1}$ a pure, entangled state is
created.  We call this first stage of the protocol pure-state
conversion measurement (PCM).  In the case of identical states, i.e.
$\alpha=\beta,\,\gamma=\delta,\,\lambda=\nu$, PCM already yields a
singlet. Otherwise the state can be transformed into a singlet via the
`Procrustean method'~\cite{Bennett96}. The total success probability
of this protocol then yields the SCP
\begin{equation}
p = 2\lambda\nu \min[ \alpha(1-\beta-\delta),\beta(1-\alpha-\gamma) ].
\label{eq:prob}
\end{equation}
Note that the SCP~(\ref{eq:prob}) corresponding to our protocol
coincides with the highest possible probability for the conversion of
two mixed states into a singlet~\cite{Jane02}.

The partition of the network into two groups $\mathcal{A}$ and
$\mathcal{B}$ is arbitrary as long as one group contains $A$ and the
other contains $B$. Hence, a singlet between $A$ and $B$ [dashed line
in fig.~\ref{fig.1}(a)] can only be established if the groups are
connected by at least two states of the form (\ref{eq:state1}), when any operation within each group is allowed.
This has to be true for {\em all} possible partitions.  Therefore a necessary condition is found for
{\em any} strategy aiming to establish a singlet between two nodes of
a mixed-state network with a finite probability: There must be at
least two distinct `paths' of edges of the form~(\ref{eq:state1})
connecting the corresponding nodes. In fig.~\ref{fig.1}(a) this is
indicated by two spatially distinct paths of bonds, i.e. in this case
it is sufficient that each bond in the path contains one edge.  Note
that the remaining qubits which are not contained in this path are
irrelevant and can therefore be in arbitrary states. Although we
allowed global operations within a group to demonstrate the necessity
of the condition, it turns out that it is also sufficient for a
singlet to be formed using LOCC. This is because entanglement swapping
transforms two states of the form~(\ref{eq:state1}) into a state of
the same form with a finite success probability. We can therefore
create a state $\rho(\alpha,\gamma,\lambda)$ between two nodes of the
network given that these nodes are connected by a path consisting of
states of the same type. Two such states, originating from two paths,
are then converted into a singlet by the protocol described above.
Unfortunately, this scheme leads in general to an
exponential decrease of entanglement fidelity~\cite{Duer99}, and thus
success probability, with the number of swapping operations and is
therefore impractical.  In the following we therefore specialize on
particular networks which allow for efficient protocols.

The condition described above is fulfilled by regular networks as
shown in fig.~\ref{fig.1}(b) in which each node is connected to its
nearest neighbors by a bond which consists of multiple edges.  We
assume that each bond is identical but allow for different edges
within a single bond.  This setup can be generalized to arbitrary
geometries, e.g. square or honeycomb, including lattices in
higher dimensions. CEP is achieved if each bond can be converted into
a singlet with a probability exceeding the percolation threshold (e.g.
$p_{th}\approx 0.347$ for a triangular lattice~\cite{book}) which is
possible if and only if each bond consists of at least two states of
the form~(\ref{eq:state1}).  For the sake of simplicity we assume in
the following that each edge is - up to local unitaries - of the form
$\rho(\alpha,\lambda)\equiv \rho(\alpha,\gamma=0,\lambda)$.  Setting
$\gamma=0$ is not a substantial restriction but keeps the following
equations manageable. All protocols in this paper can also be
performed if $\gamma\ne 0$. Note that purifiable mixed states (PMSs)
of the type $\rho(\alpha,\lambda)$ form the state of two entangled atomic
ensembles in particular quantum repeater schemes~\cite{Duan01}.

\begin{figure}[t]
  \centering\includegraphics[]{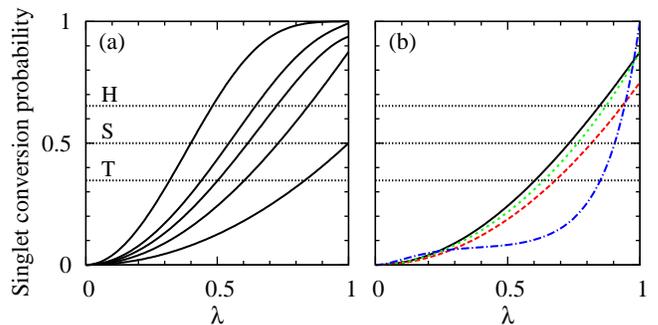}
\caption{%
  (color online) (a) Probability of generating a singlet
  using the recycling protocol for $\alpha = 1/2$ and $n=2,4,6,8,16$
  (bottom to top). (b) SCP Eq.~(\ref{Equ:GP-ProbFn}) for
  $n=3$ (red, dashed line), $n=4$ (green, dotted line) and $n=16$
  (blue, dashed-dotted line). As a reference, the $n=4$ curve from
  (a) is replotted. The percolation thresholds for triangular (T),
  square (S) and honeycomb (H) lattices are given by the horizontal
  lines.  }
\label{fig.2}
\end{figure}

If there are {\em exactly} two edges between each node the SCP is
given by Eq.~(\ref{eq:prob}).  However, allowing for more than two
edges the SCP can be greatly increased. Indeed, given $n$ identical
PMSs of the form $\rho(\alpha,\lambda)$, we developed a protocol which
makes use of `distillable subspaces'~\cite{Jane02}.  These subspaces
are constructed such that the projection of
$\rho(\alpha,\lambda)^{\otimes n}$ into the subspace is pure and
entangled. Assuming, without loss of generality, that $\alpha\ge 1/2$,
the corresponding SCP is given by
\begin{align}
  P(n,&\alpha,\lambda)=\sum_{l=0}^n  \lambda^{n-l} (1-\lambda)^l \binom{n}{l} \times  \nonumber \\
  &\left( \sum_{k=1}^{n-l-1} \frac{\alpha^{n-l-k} (1-\alpha)^k
      \binom{n-l}{k} (\binom{n-l}{k}-1)}{\binom{n}{k}-1}\right).
\label{Equ:GP-ProbFn}
\end{align}
Examples are shown in fig.~\ref{fig.2}(b). Note that for the $n=2$
case this yields the same SCP as the protocol leading to
Eq.~(\ref{eq:prob}). In most situations the SCP is inferior to the
alternative methods developed below and therefore we will not describe
this method here in detail. The derivation of
Eq.~(\ref{Equ:GP-ProbFn}) and the corresponding purification protocol
can be found elsewhere~\cite{Broadfoot09}.

The SCPs $P(n,\alpha,\lambda)$ can be significantly improved by
grouping $n$ identical PMSs into sets of $m$ and converting each of
these sets into a singlet.  For example, for $m=2$ we attempt to
convert pairs of PMSs connecting two nodes $A$ and $B$ into singlets
by using PCM. If this fails for a given pair of PMSs the
protocol generates another PMS if we measure both qubits in state
$\ket{0}$.
This PMS can successively be used for purification. More precisely,
given $n$ copies of a state $\rho(\alpha,\lambda)$ (with
$\alpha\ge1/2$) we apply the purification protocol on groups of two
and, if no singlet is obtained, repeat on the remaining PMSs.  The
coefficients for the PMSs after $k$ repetitions when no singlet is
created are given by
\begin{eqnarray}
\alpha_k &=&  \frac{\alpha_{k-1}^2}{1-2\alpha_{k-1}+2\alpha_{k-1}^2}, \\
\lambda_k &=& \frac{\lambda_{k-1}^2(1-2\alpha_{k-1}+2\alpha_{k-1}^2)}{1-2\lambda_{k-1}+2\lambda_{k-1}^2(1-\alpha_{k-1}+\alpha_{k-1}^2)   }  ,
\end{eqnarray}
where $\alpha_0=\alpha$ and $\lambda_0=\lambda$.  For states of the
form $\rho(\alpha_k,\lambda_k)\otimes\rho(\alpha_k,\lambda_k)$ the
probability of obtaining a PMS is $c_k=1-2\lambda_k +
2(1-\alpha_k+\alpha_k^2)\lambda_k^2$. Furthermore, the probability
of a singlet purification step failing without obtaining a PMS,
i.e.~measuring the two qubits in different states, is given by
$f_k=2 \lambda_k (1-\lambda_k)$. The probability of failing when we
use this {\em recycling protocol} on $n$ states of the form
$\rho(\alpha_i,\lambda_i)$ is then found to be
\begin{equation}
F_n(i) = \sum_{k=0}^{\lfloor \frac{n}{2} \rfloor} \left( \binom{\lfloor \frac{n}{2} \rfloor}{k} f_i^{\lfloor \frac{n}{2} \rfloor -k} c_i^k F_k(i+1) \right),
\label{Equ:Rec-iterativeFail}
\end{equation}
where $F_0(i) = 1$. Consequently, the probability of successfully
generating a singlet by applying the procedure to $n$ states of the
form $\rho(\alpha,\lambda)$ is $1-F_n(0)$ which is calculated
iteratively. Examples are shown in fig.~\ref{fig.2} for $\alpha=1/2$
together with the required percolation thresholds for different
lattice geometries~\cite{book}. A basic setup suitable for CEP is a
double-edged triangular lattice [see fig.\ref{fig.1}(b)]. The edges
can be converted to singlets and if the chance of this is larger than
the percolation threshold, i.e. if $2\lambda^2 \alpha(1-\alpha) >
2\sin(\pi/18) \approx 0.347$, an infinite cluster will form and a
singlet can be created between any two nodes within the cluster.
However since the singlet conversion probability never exceeds $1/2$
in this case more edges per bond are required in other geometries.
Already three edges give a SCP of $3\lambda^2\alpha(1-\alpha)\le 3/4$
[see Eq.~(\ref{Equ:GP-ProbFn})] which is sufficient for square or
honeycomb lattices for large enough $\lambda$ as shown in
fig.~\ref{fig.2}.
\begin{figure}[t]
\centering\includegraphics[width = 8.7 cm]{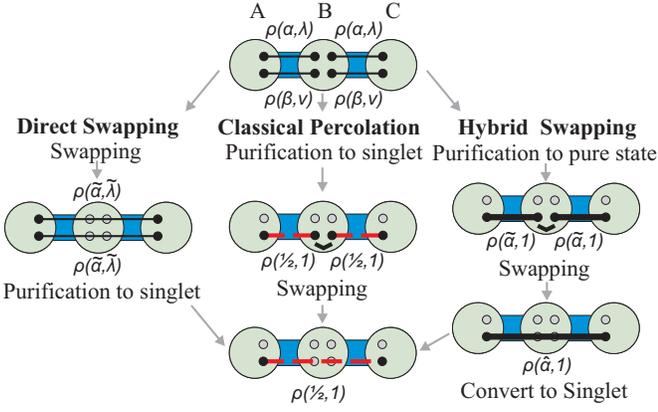}
\caption{%
    (color online) Possibilities to generate a singlet
    between two nodes $A$ and $C$ via an intermediate node $B$. The
    black, thick lines in the hybrid swapping scheme indicate pure but
    not maximally entangled states.}
\label{fig.3}
\end{figure}

\begin{figure}[t]
\centering\includegraphics[]{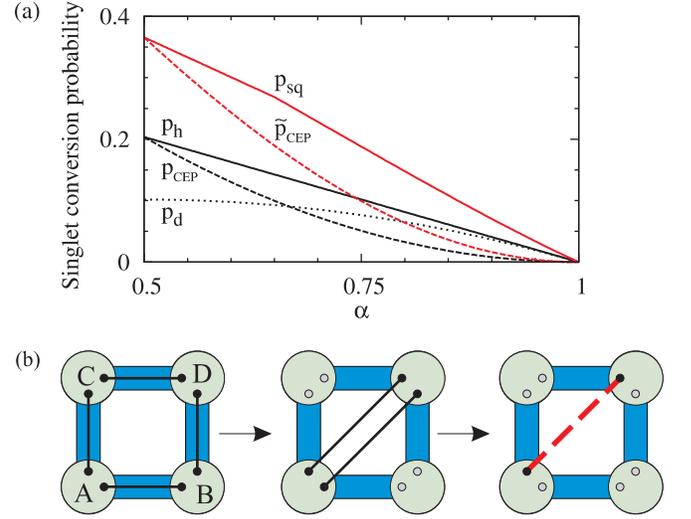}
\caption{%
    (color online) (a) Comparison of singlet conversion
    probabilities for the different strategies in the 1D configuration
    shown in fig.~\ref{fig.3} ($p_{CEP},\,p_h,\,p_d$) and in the
    square configuration ($\tilde p_{CEP},\,p_{sq}$) for
    $\lambda=\nu=0.95,\,\beta=0.5$.  (b) Square network of
    single-edged bonds. Here it is impossible to purify each
    individual bond but by using a swapping operation we create a
    purifiable double-edged bond.}
\label{fig.4}
\end{figure}

Although CEP is a promising tool for entanglement distribution it is
known that in a network of pure states the SCP can be improved by
certain quantum
strategies~\cite{Acin07,Perseguers08,Lapeyre09,cuquet09,perseguers09b}.
As we will show in the following, this is also the case in mixed-state
networks. A simple example is a protocol involving entanglement
swapping consisting of a 1D setup of three nodes $A$, $B$ and $C$ [see
fig.~\ref{fig.3}].  Partially entangled states between $A$ and $B$ and
between $B$ and $C$ are joined at $B$ leading to a singlet between $A$
and $C$.  Since the nodes have to be connected by at least two states
of the form~(\ref{eq:state1}) there are multiple possibilities to
create a singlet between nodes $A$ and $C$ as illustrated in
fig.~\ref{fig.3}.  We assume here that each bond consists of two edges
$\rho(\alpha,\lambda)$ and $\rho(\beta,\nu)$.  CEP leads to the
overall success probability
\begin{equation}
p_{CEP} = [2\lambda\nu \min(\alpha(1-\beta),\beta(1-\alpha))]^2,
\label{Equ:Pcep}
\end{equation}
see Eq.~(\ref{eq:prob}).  This
can be improved by a {\em hybrid swapping} protocol in which we first
convert each of the two bonds into a partially entangled pure state
using PCM. If $\alpha\ne\beta$ we proceed by entanglement swapping
followed by the Procrustean method (if $\alpha=\beta$ the method is
identical to CEP). The overall success probability for this protocol
is
\begin{equation}
p_h =2\lambda^2\nu^2[\alpha+\beta-2\alpha\beta]\min[\alpha(1-\beta),\beta(1-\alpha)].
\label{Equ:Ph}
\end{equation}
A third method is given by {\em direct swapping} where we first perform
entanglement swapping on the states $\rho(\alpha,\lambda)$ (from bond
$A-B$) and $\rho(\beta,\nu)$ (from bond $B-C$) and analogously on the
two remaining states (swapping on identical states yields lower SCPs).
This leads to two PMSs between $A$ and $C$.  Subsequent purification
(using PCM and the Procrustean method) yields the overall success
probability
\begin{equation}
p_{d} = 2\lambda^2\nu^2\alpha\beta(1-\alpha)(1-\beta).
\label{Equ:Pd}
\end{equation}
We see that $p_h\ge p_{CEP}$ and $p_h > p_{d}$. Furthermore,
$p_{d}>p_{CEP}$ if $2\min[\alpha(1-\beta),\beta(1-\alpha)]<\max[\alpha(1-\beta),\beta(1-\alpha)]$.
This means the hybrid protocol has the highest success probability if
$\alpha\ne\beta$.  However, if $\alpha=\beta$ CEP can not be
outperformed by direct or hybrid swapping.  The three methods are
compared in fig.~\ref{fig.4}(a): hybrid and direct swapping can lead to
substantially higher SCPs than CEP.

Although direct and hybrid swapping is used here in a 1D setup, they
can easily be embedded into a network of higher dimensions reducing
the amount of required resources or leading to higher success
probabilities. A simple example is given by a 2D square network as
shown in fig.~\ref{fig.4}(b). Direct swapping can be used to connect
the qubits in opposite nodes and subsequent purification leads to a
singlet between opposite corners. This method illustrates how two
different paths of edges can be used for singlet generation between
distant nodes. It does not require two edges per bond.  However, if we
{\em do} have two edges per bond we can use the hybrid method to
convert each bond into a pure state $\ket{\hat\alpha}\equiv
\sqrt{\hat\alpha} \ket{00} + \sqrt{1-\hat\alpha}\ket{11}$ with
\begin{equation}
\hat \alpha = \max[\alpha(1-\beta),\beta(1-\alpha)]/(\alpha+\beta-2\alpha\beta)
\label{Equ:alphahat}
\end{equation}
by using PCM which succeeds with probability $p_c =
\lambda\nu(\alpha+\beta-2\alpha\beta)$.  If this yields only two
states $\ket{\hat\alpha}$ which have a common node ($B$ or $C$),
entanglement swapping can be performed followed by the Procrustean
scheme.  If the conversion succeeds such that all four states have the
form $\ket{\hat\alpha}$ they are connected (e.g. at nodes $B$ and $C$)
via `$XZ$-entanglement swapping'~\cite{Perseguers08} leading to two
pure states (between $A$ and $D$) of the form $\ket{\tilde\alpha}$
with
\begin{equation}
\tilde\alpha= (1+\sqrt{1-16\hat\alpha^2(1-\hat\alpha)^2})/2.
\label{Equ:alphatilde}
\end{equation}
These can be distilled into a singlet with probability
$\min[1,2(1-\tilde\alpha^2)]$ by using a protocol based on
majorization theory~\cite{Nielsen01} described
in~\cite{Perseguers08,Lapeyre09}. The overall chance of succeeding in
generating a singlet is then given by
\begin{equation}
 p_{sq} = 4p_c^2(1-p_c^2)(1-\hat{\alpha}) + p_c^4\min (1, 2(1-\tilde{\alpha}^2)).
\label{Equ:SquareProb}
\end{equation}
Connecting two corners via CEP has a success probability of $\tilde
p_{CEP} = 1-(1-p_{CEP})^2$ which can be significantly smaller
than Eq.~(\ref{Equ:SquareProb}) as shown in fig.~\ref{fig.4}(a).

\begin{figure}[t]
\centering\includegraphics[width=8cm]{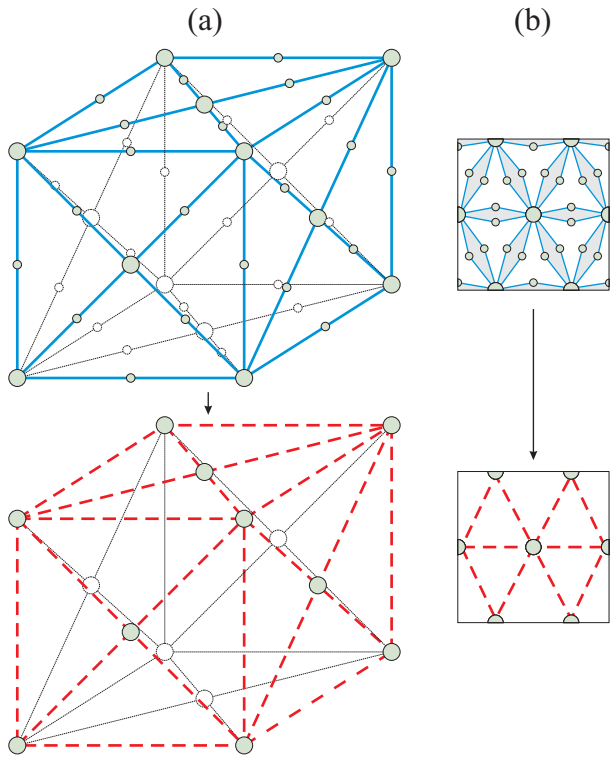}
\caption{%
    (color online) (a) Connections of a face-centred cubic
    lattice can be replaced via hybrid swapping.  Bonds (blue, solid
    lines) between nodes (small and large circles) are assumed to
    contain two edges (not shown). Hybrid swapping performed via
    intermediate nodes (smaller circles) leads to a face-centered
    lattice with singlets randomly distributed between nodes (dashed
    lines).  (b) The square protocol [cf. fig.~\ref{fig.4}(b)] is used
    in this network to achieve the percolation threshold of a
    triangular lattice. Squares are given by the shaded areas and all
    bonds are assumed to contain two edges (not shown). }
\label{fig.5}
\end{figure}

These examples of quantum strategies are only composed of a finite
number of nodes and do not exhibit a percolation threshold. However,
they can be embedded within larger networks. Two examples are
illustrated in fig.~\ref{fig.5}. We can generate singlets using small
networks like hybrid swapping and the square protocol. Doing this
results in singlets randomly distributed between neighboring nodes of
the larger network. If the probability of generating a singlet exceeds
the larger network's percolation threshold then infinite clusters will
form and enable long-distance entanglement distribution. Our quantum
strategies have a greater chance of generating a singlet compared to
CEP. This means the percolation threshold can be achieved if we use
our quantum strategies in cases where this is not possible using CEP.

For example, small swapping arrangements [see fig.~\ref{fig.3}] can be
nested within a 3D face centred cubic (fcc) lattice as shown in
fig.~\ref{fig.5}(a). The percolation threshold for a fcc lattice is
given by $p_{th} \approx 0.12$~\cite{Lorenz98}. CEP on this lattice
only creates an infinite cluster when $p_{CEP}>p_{th}$. However, the
SCP of hybrid swapping exceeds $p_{CEP}$ [see fig.~\ref{fig.4}(a)]. As
a consequence, there is a wide range of parameters $\alpha,\,\lambda$
such that $p_h>p_{th}>p_{CEP}$ and thus an infinite cluster of
singlets on the fcc lattice can only be achieved if hybrid swapping is
performed.  Similarly, the square network arrangement can be nested
within a 2D triangular lattice [see fig.~\ref{fig.5}(b)]. In this
case, the percolation threshold is satisfied by CEP if
$\tilde{p}_{CEP}>2\sin(\pi /18)$, i.e. the square SCP exceeds the
triangular lattice's percolation threshold. However, by applying the
quantum strategy on the squares we create singlets with probability
$p_{sq}>\tilde{p}_{CEP}$ which can be greater than the threshold in
cases when $\tilde{p}_{CEP}$ is smaller than the threshold [see
fig.~\ref{fig.4}(a)].

  In this paper we have found a necessary and sufficient condition for
  the generation of a perfect singlet in a quantum network where the
  nodes are initially connected by bipartite mixed qubit states. To
  form a perfect singlet between two nodes these have to be connected
  by at least two paths of states of the form~(\ref{eq:state1}). If the
  states forming the bonds in such a network are not of this type a
  singlet can not be generated with finite probability. By considering
  regular networks we have devised efficient protocols for the
  distillation of singlets and have shown that percolation can be
  applied to achieve long-distance entanglement distribution. This
  extends on previous work considering pure states~\cite{Eisert07}.

Furthermore, we have shown that it is principally possible to find
quantum strategies for entanglement distribution which outperform CEP.
The extension of CEP by such quantum strategies can lead to improved
success probabilities for singlet generation, e.g. by nesting small
structures like the discussed square setup into large-scale
networks~\cite{Acin07,Perseguers08,Lapeyre09}. However, like in the
pure-state case, the optimal strategy, i.e. the strategy with the
highest possible success probability using the least amount of
physical resources, remains unknown. Indeed, an optimal scheme for
entanglement distribution might even go beyond the restriction to
bipartite entanglement: Networks with multipartite entangled (mixed)
states might give further
advantages~\cite{perseguers09c,perseguers09b}.

Other schemes for distributing entanglement in quantum
  networks have been found in the presence of bit-flip noise
  \cite{Perseguers08b}, and Werner states allow the
  generation of long-range entanglement in a cubic network
  \cite{raussendorf05,perseguers09c}. However, although the resulting
  states can have a high fidelity they are not perfect singlets. Yet,
  the ability to purify the states to high, non-unity fidelities make
  these other strategies very interesting.

Despite this, the simplicity of CEP makes it a very
promising framework for the development of long-distance entanglement
distribution in quantum networks.

This research was supported by the the EPSRC (UK) through the QIP IRC
(GR/S82176/01) and the ESF project EuroQUAM (EPSRC grant EP/E041612/1).

\end{document}